# DFT based study on structural stability and transport properties of Sr$_3$AsN: A potential thermoelectric material

Enamul Haque, and M. Anwar Hossain

Department of Physics, Mawlana Bhashani Science and Technology University

Santosh, Tangail-1902, Bangladesh

Email: enamul.phy15@yahoo.com

**Abstract**

Antiperovskite materials are well known for their high thermoelectric performance and gained huge research interest. Here, we report the structural stability and transport properties of Sr$_3$AsN from a precise first-principles study. The calculated equilibrium lattice parameters are in a good agreement with the available data. We find that Sr$_3$AsN is a mechanically, energetically and dynamically stable at ambient condition. Our calculated electronic structure indicates that it is a direct bandgap semiconductor, with a value ~1.2 eV. Sr-4d and N-2p orbitals mainly formulate the direct bandgap. This antiperovskite possesses a high Seebeck coefficient. Although its lattice thermal conductivity is comparatively low, electronic thermal conductivity is very high. The calculated maximum TE figure of merit is 0.75 at 700 K, indicating that it is a potential material for thermoelectric applications.

## 1. Introduction

The search for new high-performance thermoelectric materials has been attracted to the researcher much because the high-performance thermoelectric material is very useful in practical energy conversion applications [1]. The thermoelectric performance of a material can be characterized by a dimensionless quantity, called a figure of merit, ZT, defined as [2,3] ZT=$\frac{S^2\sigma T}{k}$, where the quantities $S$, $\sigma$, $k = k_e + k_l$ and $T$, stand for Seebeck coefficient, electrical conductivity, thermal conductivity (sum of the electronic and lattice contribution) and imposed temperature, respectively. Generally, semiconductors have a high Seebeck coefficient, low thermal conductivity. Antiperovskites materials, such as $Sr_3SbN$, is a semiconductor, with a bandgap ~1.15 eV and possesses high thermopower and power factor [4,5]. Most of the high performance thermoelectric materials have a bandgap around 0.5-1.5 eV [1,6–8]. Therefore, its isotypic compound may have a bandgap within this range and hence, high performance thermoelectric conversion efficiency.

$Sr_3AsN$, one such a compound, was predicted by Beznosikov from the analysis of chemical bonding and admissible atomic radius [9]. Recently, some authors have reported its elastic, electronic and optical properties, without considering its energetic and dynamic stability [10–12]. The dynamical stability is an essential criterion for a crystal to be synthesized successfully in the laboratory [13]. Some compounds have not been possible to synthesize due to their dynamical instability [13,14]. The electronic properties of $Sr_3AsN$ have been reported in most of the studies by using PBE functional [10,11]. Since PBE functional underestimate bandgap about 50% of the experimental value [15], the reported bandgap value is ~0.5-0.8 [10,11]. From the above circumstances, it is very interesting to study the structural stability, precise electronic structure and transport properties of $Sr_3AsN$.

In the present work, we study the structural stability from lattice dynamics, energetic stability from formation energy and free energy calculation, precise electronic structure by using PBE and different version of BJ potential [16], electron and phonon transport properties. Our analysis reveals that the studied compound is mechanical, energetically, and dynamically stable and we hope that it can be synthesized successfully in the laboratory. The calculated direct bandgap value is now more precise and it is 1.2 eV. Although lattice thermal conductivity is low, the electronic thermal conductivity is very high, resulting in the suppression of thermoelectric performance. It may be possible to reduce electronic thermal conductivity by co-doping with the suitable element.

## 2. Computational methodology

The structural optimization was performed by using a full potential linearized augmented plane wave (FP-LAPW) with generalized gradient approximation of Perdew-Burke- Ernzerhof (GGA-PBE) [17,18], as implemented in WIEN2k [19]. We used $10 \times 10 \times 10$ k-point and kinetic energy cutoff of RKmax=7.0. We modeled the muffin tin sphere with radii 2.37, 2.5, 2.26 Bohr for Sr, As, and N, respectively. By using optimized structure, we performed elastic and optical properties in WIEN2k. For electronic structure, we used $21 \times 21 \times 21$ k-point and different versions of BJ potential. We performed the self-consistent field calculation (SCF) again by using a finer $54 \times 54 \times 54$ k-point mesh to generate the necessary inputs (energy eigenvalues, Fermi energy) to solve Boltzmann transport equation, as implemented in BoltzTrap [20]. The lattice thermal conductivity has been calculated by using the finite displacement approach [21,22], displacing each atom ( in $2 \times 2 \times 2$ supercell) by 0.06 Å simultaneously, which is implemented in Phono3py [23]. We have used a mesh of $4 \times 4 \times 4$ k-point to calculate the required harmonic (second-order forces) and anharmonic (third order forces) force calculation in Quantum espresso package [24] using GGA-PBE functional, ultrasoft pseudopotential. The BZ integration in the q-space using a $21 \times 21 \times 21$ q-point mesh has been performed and lattice thermal conductivity has been calculated by using the equation $\kappa = \frac{1}{NV}\sum_\lambda C_\lambda \boldsymbol{v}_\lambda \otimes \boldsymbol{v}_\lambda \tau$, where V stands for the volume of the unit cell, $\boldsymbol{v}$ for the group velocity, $\tau$ for the SMRT for the phonon mode λ, and $C_\lambda$ for mode dependent phonon heat capacity. The Phono3py program has been found to be successful in the prediction of the lattice thermal conductivity of some materials [25–29].

## 3. Results and discussions

The antiperovskite Sr$_3$AsN crystallizes in a cubic structure of space group $Pm\overline{3}m$ (#221). The three Sr-atoms occupy 3c (0, ½, ½), one As and one N-atom occupy 1a (0, 0, 0), and 1b (½, ½, ½) Wyckoff positions, respectively. These three Sr-atoms are bonded with N, while no bond exist beween Sr and Sn atoms.

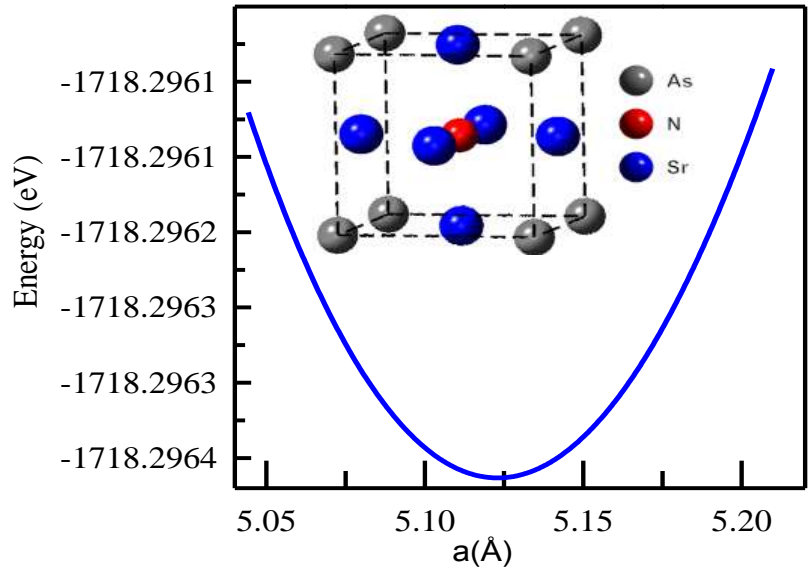

Fig. 1: The calculated total energy of Sr$_3$AsN as a function of lattice parameter. The inset figure shows the ground state structure of the studied compound.

The equilibrium crystal structure and ground state energy as a function of lattice parameters are shown in Fig. 1. From energy minimization (solving the equation of state (EOS), our calculated equilibrium lattice parameter is listed in Table-1, with other theoretical available data.

Table-1: Calculated lattice parameter with other theoretical available data.

|  | PBE(this) | PBE(Ref. [12]) | PBE(Ref. [10]) | PBE(Ref. [9]) | WC(Ref. [12]) |
|---|---|---|---|---|---|
| a(Å) | 5.1266 | 5.1226 | 5.118 | 5.03-5.084 | 5.0526 |

We see that the calculated lattice parameter is in a good agreement with the available data. We will use the equilibrium lattice parameters in the subsequent calculations.

### 3.1. Structural stability

Before experimentalists try to synthesize a known compound, it is the best way to check its theoretical stability. First, we consider the energetic stability of $Sr_3AsN$. This criterion can be check from the formation energy of the compound. The formation energy of a compound can be calculated by taking energy difference between the total energy of the compound and sum of the total energy of the individual constituents. Our calculated formation energy is -1.51 eV/Sr.

Table-2: Calculated Elastic constants of $Sr_3AsN$, with other available data.

| cij | Present (GPa) |  | PW(GPa) [10] |  | PBE(GPa) [11] |  |
|---|---|---|---|---|---|---|
| $c_{11}$ | 95 | B | 110 | B | 98 | B |
| $c_{12}$ | 23 | 47 | 17 | 50 | 15 | 42 |
| $C_{44}$ | 29 |  | 35 |  | 46 |  |

The negative formation energy indicates that the studied compound is probable to form energetically.

Another way, we can check the elastic stability from calculated elastic constants. Our calculated elastic constants fairly agree with the available theoretical values. The slight deviation is due to the different structural relaxation used in their calculations.

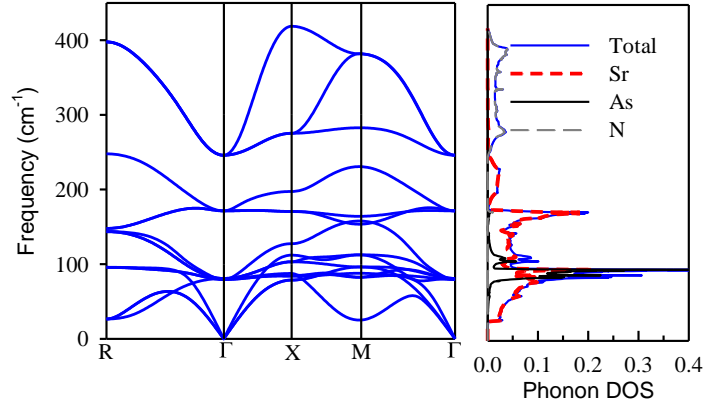

Fig. 2: Calculated phonon dispersion relations of Sr3AsN along high symmetry k-point (Left panel). The right panel shows the total and atom projected phonon density of states.

The necessary and sufficient conditions for a crystal to be elastically stable are:

$$C_{11} - C_{12} > 0 \;;\; C_{11} + 2C_{12} > 0 \;;\; C_{44} > 0$$

From the Table-2, it is clear that the calculated elastic constants of $Sr_3AsN$ fulfill these criteria and hence, the studied compound is elastically stable.

The most important criterion for a crystal to be synthesized in the laboratory, it should be dynamically stable. Phonons, called the normal mode of quantum vibrations, characterize the dynamical stability of a crystal. To be a dynamical stable, the phonon frequency of the compound must be a real quantity, i.e., the structure must not have any unstable phonon branch [30,31]. The calculated phonon dispersion and phonon density of states are presented in Fig.2. The acoustic modes mainly arise from Sr-atoms, and the peak around 100 cm-1 arises from As-atom. N-atom gives rise to all the optical modes. From Fig. 2, it is clear that $Sr_3AsN$ in cubic phase does not have any unstable phonon modes, and thus, it is a dynamically stable phase. From the above analysis, we predict that $Sr_3AsN$ in the cubic phase is highly probable to form in the laboratory. Since $Sr_3AsN$ fulfills all the stability criteria, we can proceed to the next section to study its transport properties and find its potential applications.

## 3.2. Transport properties

Since the figure of merit (ZT) is inversely proportional to the thermal conductivity, i.e., sum of the electronic and lattice thermal conductivity, materials with low lattice thermal conductivity show a high thermoelectric figure of merit. First, we insight into the phonon scattering mechanism in $Sr_3AsN$. To describe this, we calculate phonon lifetime and group velocity and present in Fig. 3. In the figure of group velocity, the acoustic and optical modes of phonons are indicated by the red star and blue circle respectively.

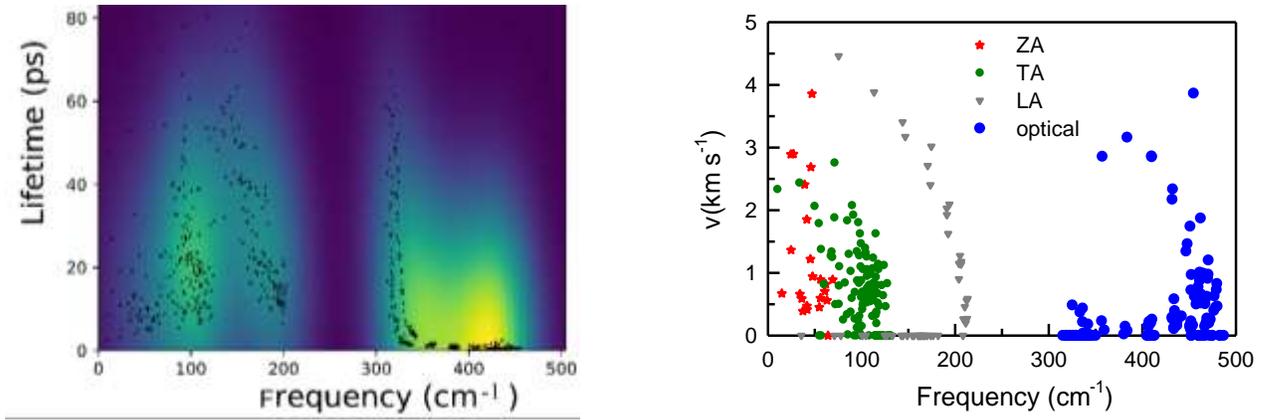

Fig. 3: Phonon lifetime and group velocity as a function of frequency.

The maximum group velocity is around 4.5 Kms$^{-1}$. The strong phonon scattering induces a short phonon lifetime. From our calculated phonon lifetime, it is clear that the phonon lifetime is relatively shorter that indicates strong phonon scattering in Sr$_3$AsN. The frequency dependent cumulative lattice thermal conductivity at 300 K is presented in the right panel of Fig. 5. We see that most of the heat is transported by acoustic phonons, below 200 cm$^{-1}$. The heat transport due to optical phonons is very small. From this point of view, we may increase the scattering channels between acoustic and optical phonons by doping/co-doping of suitable elements or nano-structuring to reduce lattice thermal conductivity further. However, we are not interested in it in the present work.

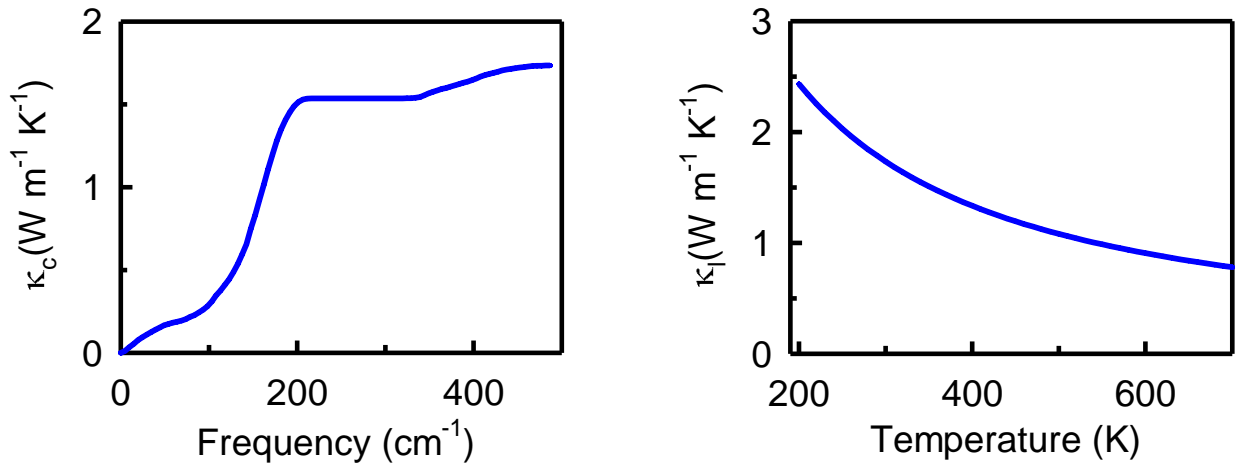

Fig. 5: Calculated: (a) cumulative lattice thermal conductivity as a function of frequency at 300 K, and (b) lattice thermal conductivity as a function of temperature.

Our calculated lattice thermal conductivity as a function of temperature is shown in the right panel of Fig. 5. We see that the lattice thermal conductivity decreases with temperature because phonon scattering increases with temperature. The calculated lattice thermal conductivity at 300 K is 1.73

W m$^{-1}$K$^{-1}$, which is much smaller than that of a typical thermoelectric material CoSb$_3$ (11.5 Wm$^{-1}$ K$^{-1}$) [32]. Such a low value of lattice thermal conductivity arises mainly from phonon scattering between the acoustic modes of Sr and As (See the calculated atom projected phonon density of states).

Now we are interested to describe electron transport in Sr$_3$AsN. Since electron transport properties are closely related to the electronic structure of the material, we will briefly review it. Our calculated electronic band structure of Sr$_3$AsN by using most accurate TB-mBJ potential is presented in the left panel of Fig. 6. The right panel shows the calculated total and projected density of states. We see that CBM and VBM are at the $\Gamma$-point, with an energy gap 1.2 eV.

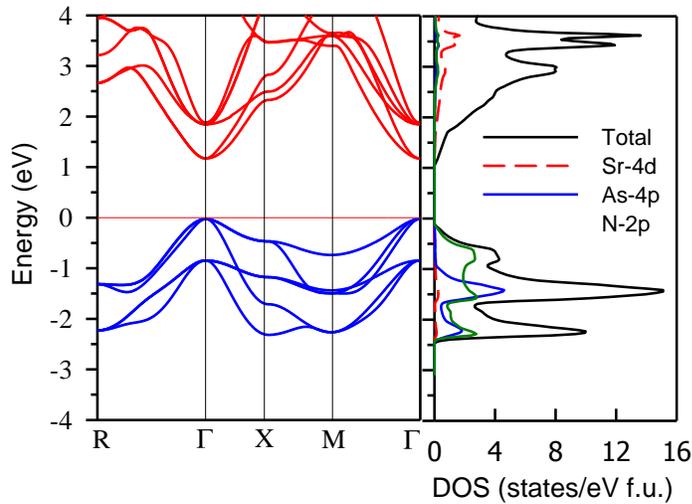

Fig. 6: Calculated electronic band structure by using TB-mBJ potential (left panel), total and atom projected density of states (right panel). The horizontal line at zero energy represents the Fermi level.

Therefore, Sr$_3$AsN is a direct bandgap semiconductor. Although the conduction bands are highly dispersive, non-flat, some valence bands are non-dispersive. We have also calculated electronic bandgap by using PBE functional and different versions of Becke-Johnson (BJ) potential [16], Tran and Blaha-modified BJ potential (TB-mBJ) [33]; new Koller, Tran, and Blaha-modified BJ potential (nKTB-mBJ) [34]; Koller, Tran, and Blaha-modified BJ potential (KTB-mBJ) for bandgap up to 7 eV [34]. Our calculated values of the bandgap of $Sr_3AsN$ are listed in Table-2.

Table-2: Calculated energy bandgap of $Sr_3AsN$ by using different functional and others available theoretical data.

| Approach | Bandgap (eV) | References |
| --- | --- | --- |
| PBE | 0.52 | This |
| PBE | 0.49, 0.30 | [10,11] |
| PBE-EV | 0.84 | [10] |
| TB-mBJ | 1.2 | This |
| KTB-mBJ | 1.22 | This |
| nKTB-mBJ | 1.24 | This |
| BJ | 0.84 | This |
| TB-mBJ(Other) | 1.12 | [12] |

From the above table, it is clear that PBE functional underestimate the TB-mBJ value by 56%. Our calculated bandgap using PBE function is in a good agreement with the available PBE based data.

It is interesting that BJ potential and Engel-Vosko GGA functional [35] give the same result. TB-mBJ and its other two modified versions also give the almost same value of bandgap. Our calculated bandgap by using TB-mBJ is slightly larger than that of reported in the Ref. [12] due to different structural relaxation used in the calculation. Since TB-mBJ usually produces an accurate bandgap than by other approaches, we have used it in the transport properties calculations.

A good thermoelectric material usually has a high Seebeck coefficient and hence, high power factor. Another important fact that it must have a low thermal conductivity. The lattice thermal conductivity of the studied compound is comparatively low and hence, it is interesting to see what a value of the figure of merit (ZT) results from such a low lattice thermal conductivity.

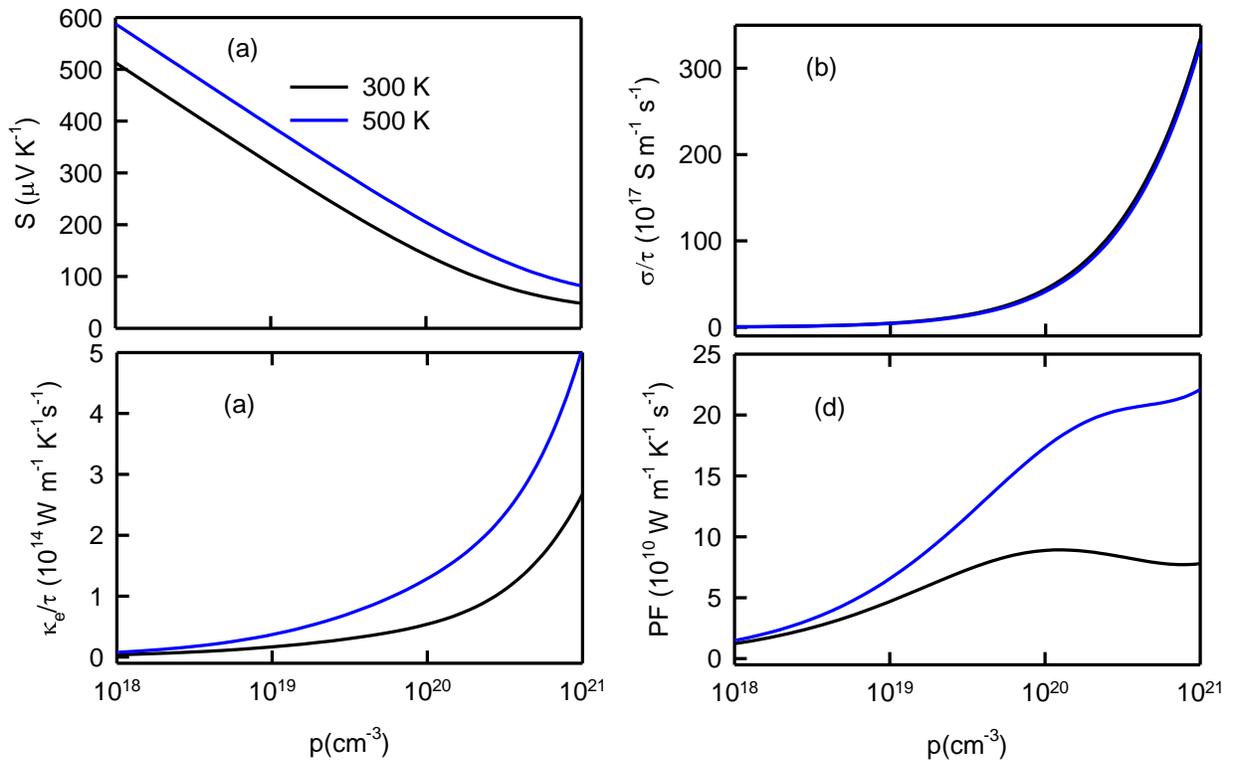

Fig. 7: Calculated: (a) Seebeck coefficient, (b) electrical conductivity, (c) electronic part of the thermal conductivity, and (d) power factor, as a function of the carrier concentration of p-type $Sr_3AsN$.

Our calculated transport parameters of p-type $Sr_3AsN$ as a function of carrier concentration are presented in Fig. 7. We see that the Seebeck coefficient decreases with the increase of carrier concentration, but it increases with temperature. The electrical conductivity ($\sigma/\tau$) electronic part of the thermal conductivity increase with carrier density, as usually. Since electrical conductivity increases with carrier density, the power factor also increases with it (at 300 K, PF increases with the carrier density up to $10^{20}$ cm$^{-3}$).

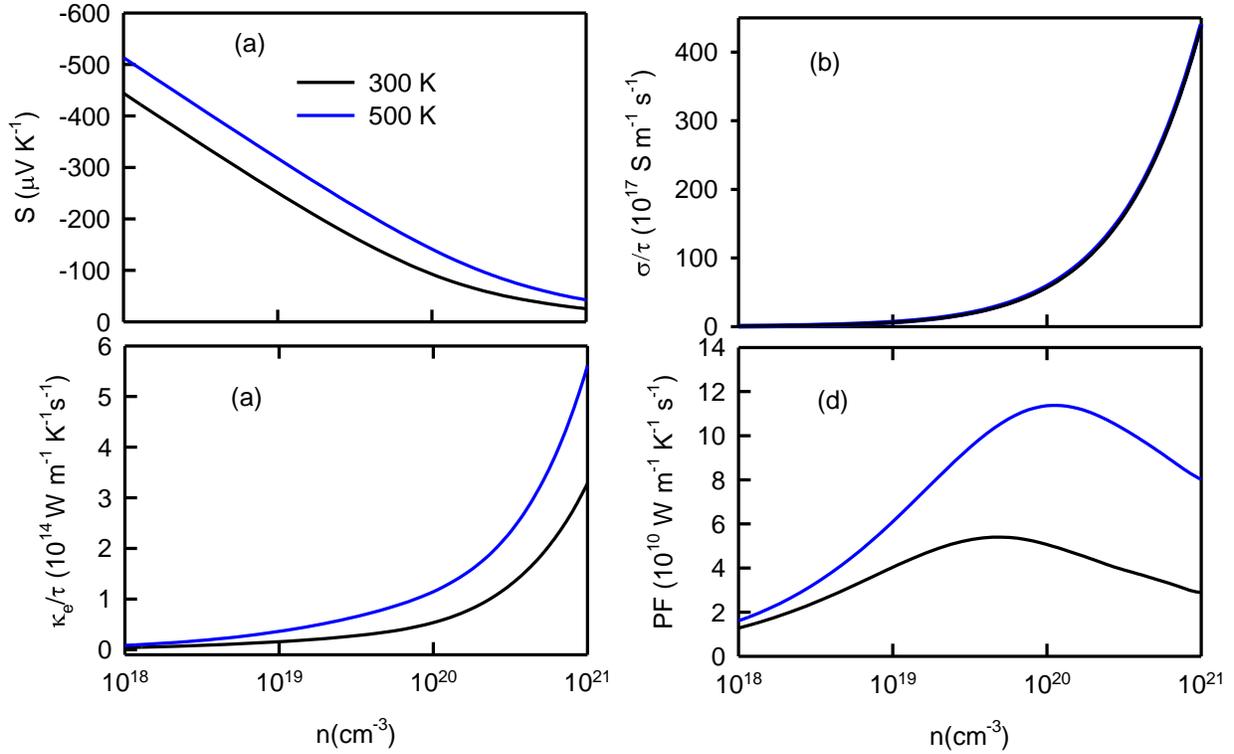

Fig. 8: Calculated: (a) Seebeck coefficient, (b) electrical conductivity, (c) electronic part of the thermal conductivity, and (d) power factor, as a function of the carrier concentration of n-type Sr$_3$AsN.

To asses which type of material is more favorable for thermoelectric energy conversion, calculated transport parameters of n-type Sr$_3$AsN as a function of carrier concentration are presented in Fig. 8. All parameters show the same trend with carrier concentration, but the Seebeck coefficient is much smaller than that of p-type while electrical conductivity and electronic part of the thermal conductivity higher than that of p-type. However, power factor (PF) of p-type $Sr_3AsN$ is larger than that of n-type. Therefore, p-type $Sr_3AsN$ is more favorable for thermoelectric energy conversion. For further description of the thermoelectric properties of $Sr_3AsN$, we will consider transport properties of holes only (p-type).

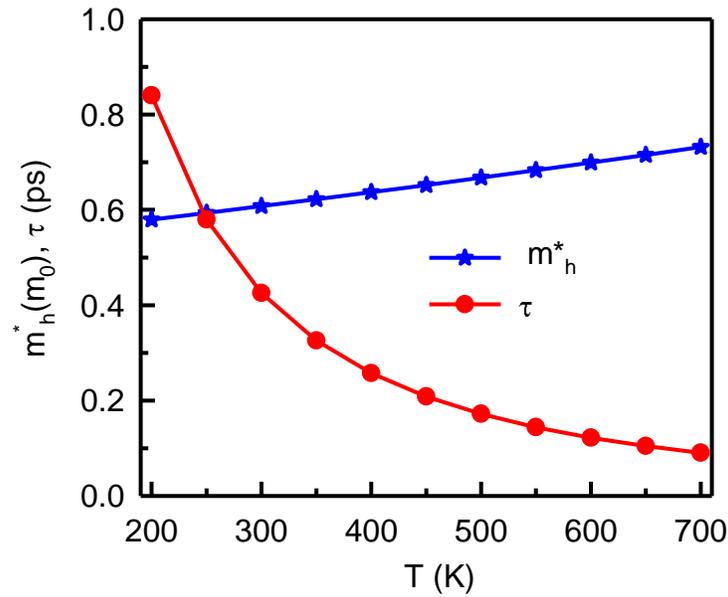

Fig. 9: Calculated transport effective mass and carrier relaxation time of $Sr_3AsN$ as a function of temperature.

As above, we have described the electrical conductivity and electronic part of thermal conductivity within the constant relaxation time approximation. To extract real values of them, we need to calculate carrier relaxation time. For this, we need to consider that acoustic phonons scatter the conduction holes, others type of scatterings are negligibly small. Deformation potential theory provides an approach to calculate relaxation time based on the equation [36]

$$\tau_{acps} = \left(\frac{2}{m^*k_BT}\right)^{3/2} \frac{\pi^{1/2}\hbar^4 c_{ii}}{3D_{ac}^2}$$

where $c_{ii}$ represents the symbol of the direction-dependent elastic constant and $D_{ac}$, the deformation potential defined as $D_{ac} = a_0 \frac{\partial E}{\partial a}$. We can calculate deformation potential constant by using finite difthe ference method and calculating the change of band energy (E) with the change of lattice parameters (a) by the applied strain. The band energy (E) has been calculated as the change of CBM energy and VBM energy due to the applied strain. Readers are suggested to consult with these references for more detail descriptions [37–39]. The calculation of direction dependent elastic constants is well described in the Ref. [36]. In this calculation, we will use the elastic constant in the direction of the pure longitudinal wave (100). The calculated deformation potential constant is 7.91 eV. Another quantity required to calculate relaxation time is effective mass (m*). For this, we have calculated transport effective mass (calculated using conductivity tensor of BoltzTrap output with the aid of transM code [40]) as a function of temperature, shown in Fig. 9. The effective mass increases with temperature due to the increase of carrier density. By using these parameters, our calculated carrier relaxation time at different a temperature is presented in Fig. 9. The carrier relaxation time is comparatively large, which indicates small holes scattering in $Sr_3AsN$. Now, we can calculate the electrical conductivity and electronic part of the thermal conductivity by multiplying the above relaxation time of the corresponding temperature. For the present purpose, we consider transport parameters at the chemical potential of 0 K Fermi energy. The temperature dependent Seebeck coefficient, electrical conductivity, total thermal conductivity, and power factor are shown in Fig. 10. We see that low temperature Seebeck coefficient is very high and it decreases with temperature. Because temperature reduces the bandgap shifting the

Fermi level at the middle of the bandgap and the intrinsic activation energy increases with temperature up to the bandgap energy.

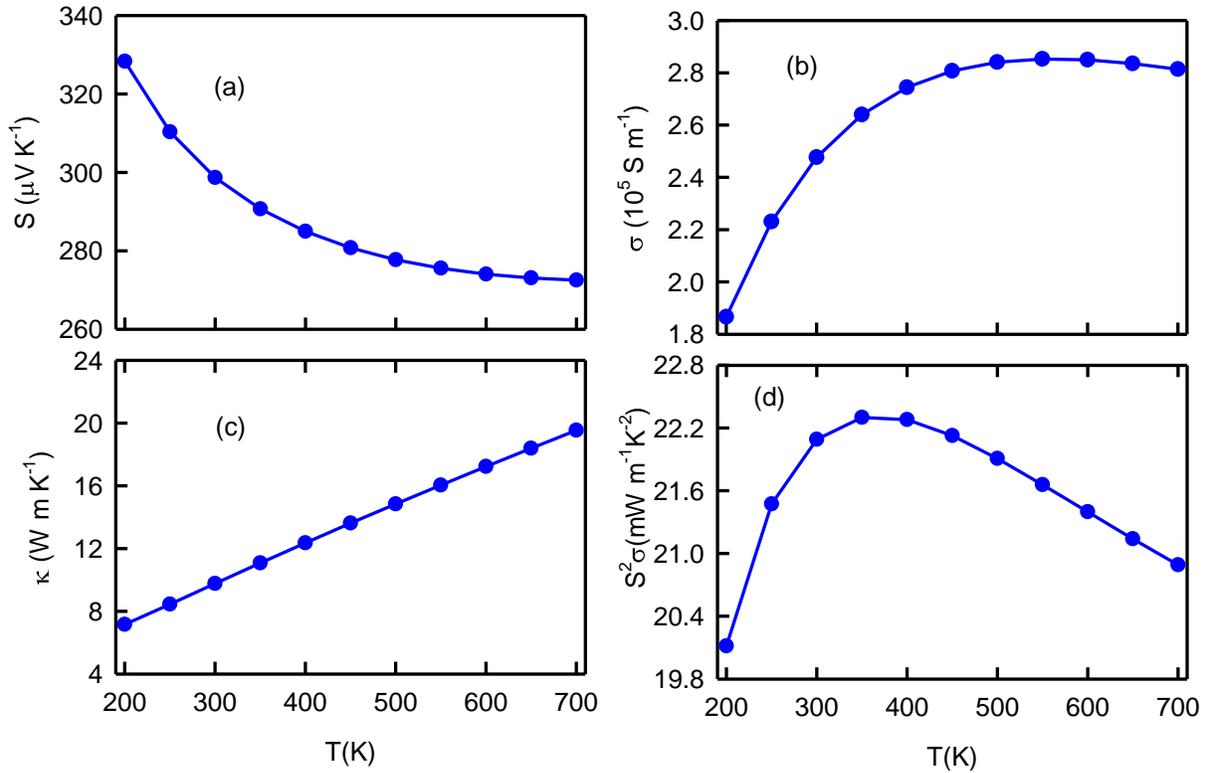

Fig. 10: Temperature-dependent thermoelectric parameters: (a) Seebeck coefficient, (b) electrical conductivity, (c) total thermal conductivity, and (d) power factor (PF).

The electrical conductivity and total thermal conductivity increase with temperature, as carrier density increases. This demonstrates the semiconducting nature of the studied compound. Above 600 K, electrical conductivity decreases with temperature indicating that $Sr_3AsN$ may become metallic above 600 K, as shown in Fig. 10(b). We obtain a maximum value of power factor 12.2 mW m$^{-1}$ K$^{-2}$ at 350 K. To reveal its applicability in thermoelectric devices, we have calculated the figure of merit (ZT) as a function of temperature, as shown in Fig. 11.

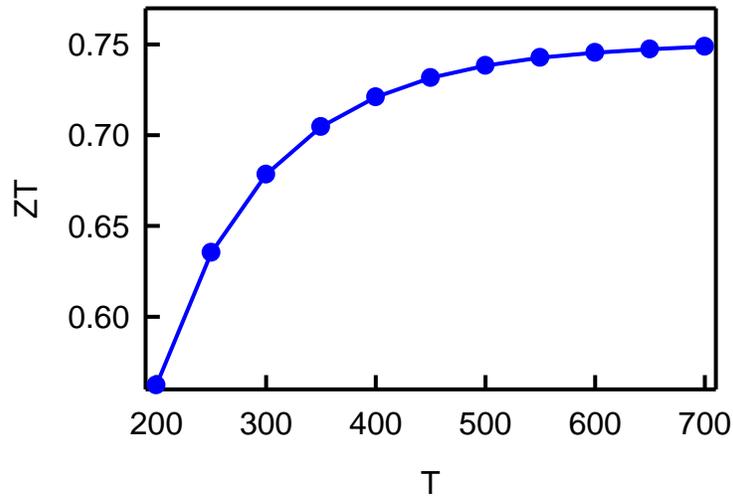

Fig. 11: Thermoelectric figure of merit of $Sr_3AsN$ as a function of temperature.

We see that ZT increases with temperature up to the studied temperature range. We obtain a maximum value 0.75 of ZT at 700 K. Therefore, $Sr_3AsN$ is a potential thermoelectric material and will be suitable for thermoelectric device applications if we can increase ZT further by alloying or nano-structuring mechanism. Because these mechanisms can reduce thermal conductivity drastically improving the scattering of electrons and phonons. Although this value of ZT is smaller than that required for practical applications, we hope that our study will inspire experimentalists to synthesize $Sr_3AnN$ and to improve thermoelectric performance by alloying with a suitable element or nanostructuring.

## 4. Conclusions

In summary, we have predicted the structural stability of $Sr_3AsN$, to reveal its possibility of synthesis, and thermoelectric transport properties to disclose its applicability in thermoelectric devices, from first-principles calculations. Our calculated equilibrium lattice parameters are in a good agreement with the available data. Our analysis on structural stability reveals that $Sr_3AsN$ is a mechanically, energetically, and dynamically stable at ambient condition. We have found from electronic structure calculations that it is a direct bandgap semiconductor, with a value ~1.2 eV. Sr-4d and N-2p orbitals mainly formulate the direct bandgap. This antiperovskite has a high Seebeck coefficient and high power factor. Although its lattice thermal conductivity is comparatively low, electronic thermal conductivity is very high. The calculated maximum TE figure of merit is 0.75 at 700 K, indicating that it is a potential material for thermoelectric applications. Although this value of ZT is smaller than that required for practical applications, we hope that further study under consideration of alloying with a suitable element or nanostructuring may improve ZT and it may be a practically applicable TE material.